\documentclass[12pt]{iopart}

\usepackage{iopams}  
\usepackage{graphicx}
\usepackage{dcolumn}
\usepackage{bm}
\begin{document}

\title[3D woodpile structure tunable plasma photonic crystal]{3D woodpile structure tunable plasma photonic crystal}

\author{B.Wang, J.A. Rodr\'iguez, and M.A. Cappelli}%
\address{
Department of Mechanical Engineering, Stanford University, Stanford, CA 94305-3032
}
\ead{cap@stanford.edu}

\begin{abstract}
A 3D woodpile structure tunable plasma photonic crystal is designed, simulated, and experimentally characterized over the S - X band of the electromagnetic spectrum. The measurements confirm that the electromagnetic response is rich in dynamics. The photonic crystal's reconfigurability, achieved through individual discharge control of the properties of the woodpile plasma columns, offers an unprecedented opportunity to better resolve the interactions between both the Bragg and localized surface plasmon modes which dominate the spectrum at lower frequencies. Both the experiments and simulations reveal evidence for the coupling of Bragg and surface plasmon modes through a Fano resonance. 

\end{abstract}


\maketitle
Photonic crystals (PCs) are artificially-ordered materials designed to transmit or reflect electromagnetic (EM) waves over a limited range of frequencies as a result of destructive and constructive Bragg scattering interferences within the periodic structure \cite{Joannopoulos2008a}. A plasma photonic crystal (PPC) \cite{HOJO2004,Sakai2012,Sakai2007a,Wang2016,Wang2015a} is a PC that uses controllable gaseous plasma elements to offer a degree of tunability or reconfigurability. The simplest PPCs consist of one-dimensional (1D) striated plasma-vacuum or plasma-dielectric layers \cite{HOJO2004,wang2018gaseous}. The most commonly-studied 1D PPCs are laser-produced plasma Bragg gratings\cite{shi2011generation,liu2017filamentary}. Two-dimensional (2D) and three-dimensional (3D) PPCs are arrays of stand-alone plasma structures \cite{Sakai2007a,Wang2016} or dielectric/metallic PCs that incorporate repeating plasma structures. As in passive PCs, PPCs can be functionalized by introducing defects that break the crystal periodicity. For example, reconfigurable line defects in a 2D PPC have been shown to result in switchable waveguiding \cite{Wang2016a}. Passive dielectric or metal PCs can also be functionalized by incorporating plasma defects to produce novel devices such as tunable bandpass filters \cite{Wang2015a} and power limiters \cite{Gregorio2017}. Plasmonic effects in PPCs, such as the excitation of bulk or localized (surface) plasmons, can further enhance the EM wave interactions\cite{wang2018gaseous, righetti2018enhanced}, resulting in deep and wide attenuation bands. In 2D PPCs, these plasmonic effects arise for TE polarization of the incident field (E-field $\perp$ to plasma rods) at frequencies below the plasma frequency, $\omega_p$. 

In this letter we describe the construction and performance of a 3D PPC consisting of a woodpile lattice structure formed from intertwined orthogonal nested 2D arrays constructed from gaseous plasma columns. The woodpile lattice structure allows for the excitation of both surface plasmon and Bragg scattering modes for arbitrary incident polarization. The plasma column properties can be individually controlled allowing reconfigurability of the PC. For example, we can turn off the plasma of all of the columns oriented in one direction, transforming the structure into a lower dimensionality 2D array. As described below, we show that there is a synergistic interaction of the two nested 2D arrays, affected somewhat by the excitation of surface plasmon polaritons. 

\begin{figure}[htbp]
\centering
\includegraphics[width=\linewidth]{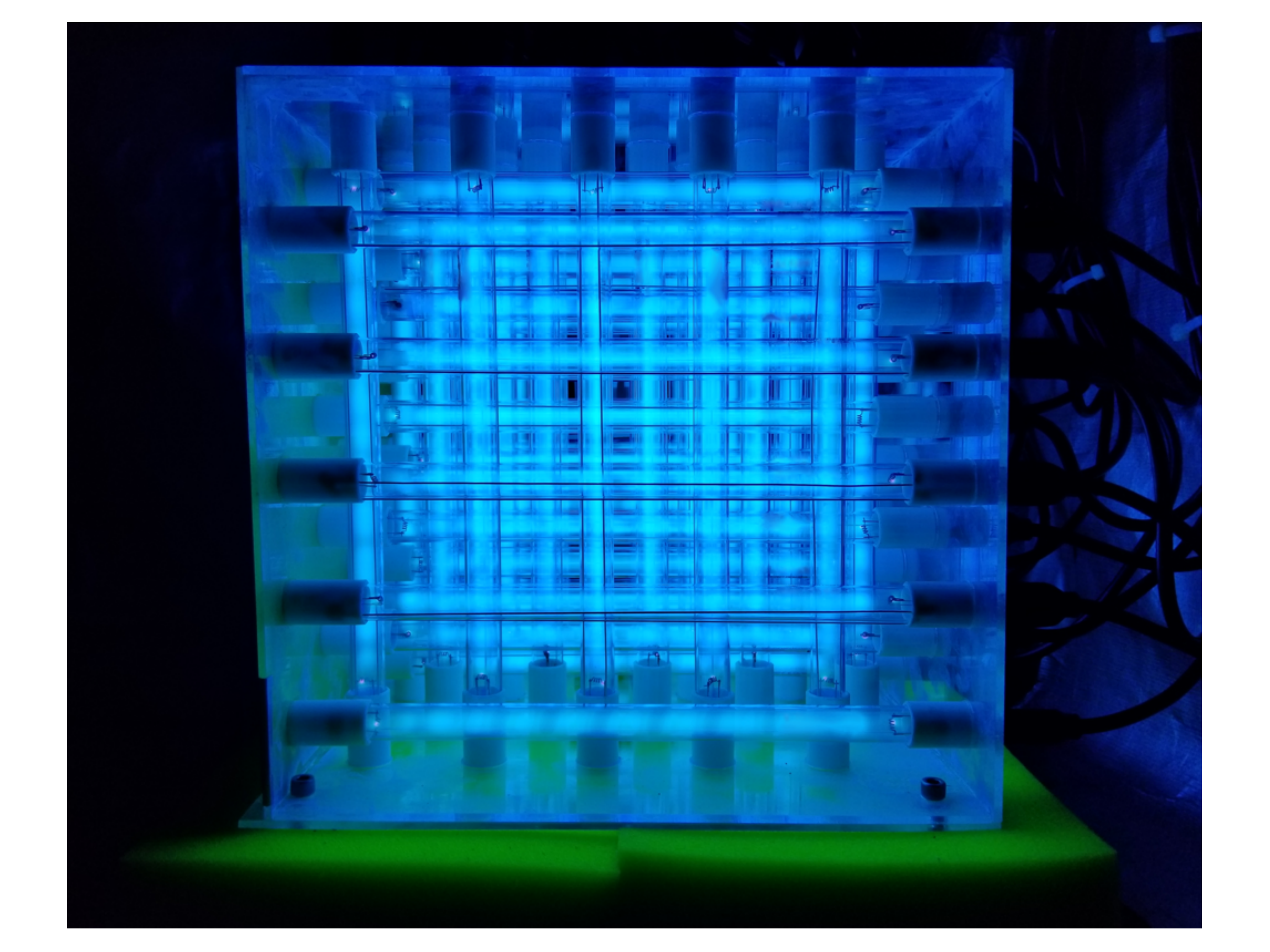}
\caption{3D plasma photonic crystal consisting of 50 plasma discharge tubes arranged in a woodpile configuration.}
\label{fig:1}
\end{figure}

\begin{figure}[htbp]
\centering
\includegraphics[width=\linewidth]{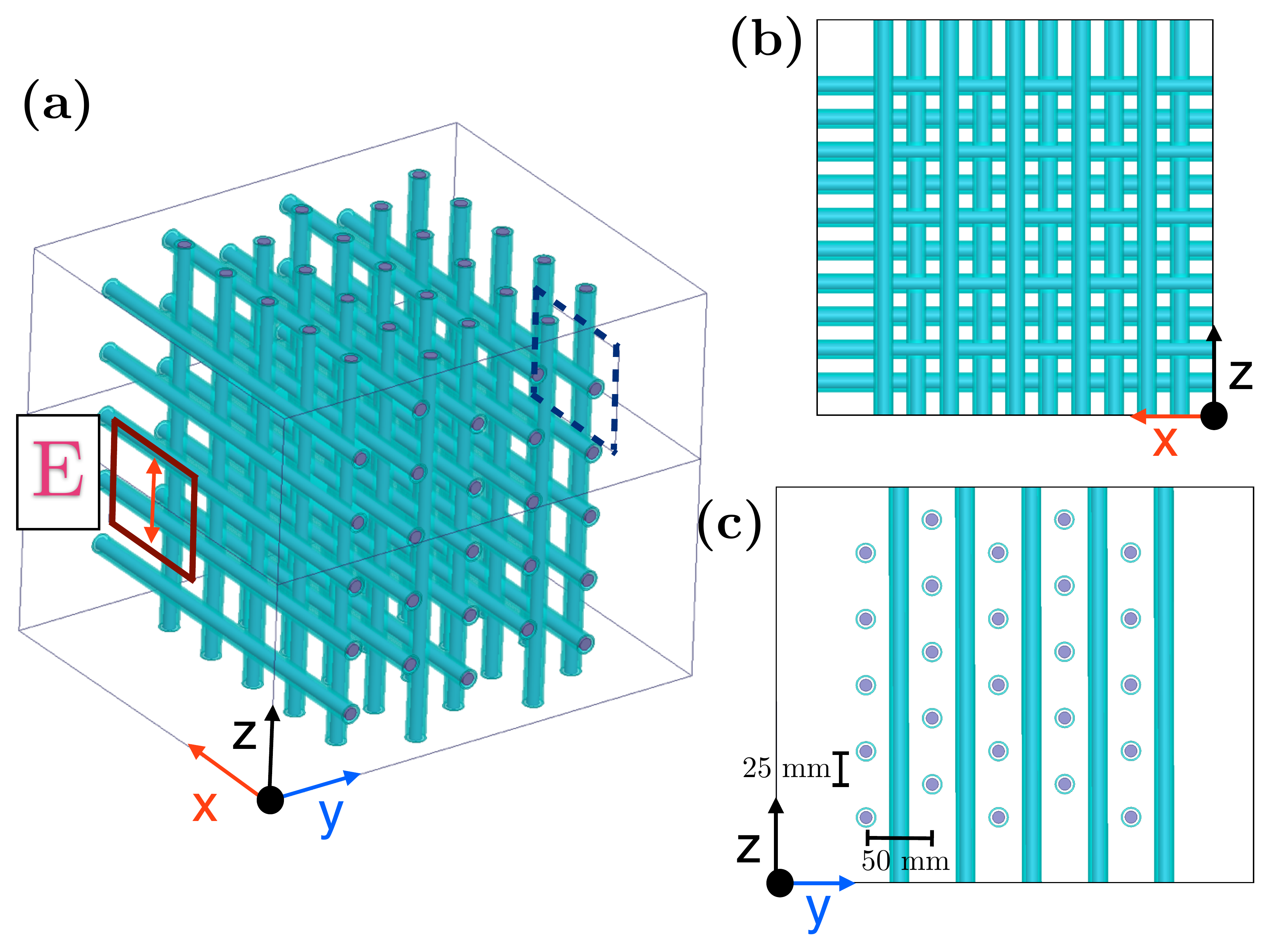}
\caption{(a) Perspective schematic of the 3D woodpile PPC. The solid red and dashed black rectangles indicate the position of the microwave horn antennas for both the experiments and simulations. (b) \textit{x-z} view of the woodpile structure showing the 3D lattice structure. The k-vector of the incident wave is in the $\textit{y}$-direction. (c) \textit{y-z} view of the woodpile structure depicting the lattice constant $\textit{a}$ = 50 mm, staggered by $\textit{a}$/2 (25 mm) in the  $\textit{z}$-direction for each array element in the  $\textit{y}$ axis. The quartz tubes are represented in teal, while the plasma rods are shown in purple.}
\label{fig:2}
\end{figure}

Figure \ref{fig:1} shows a photograph of the 3D PPC woodpile structure consisting of layers of 5 x 5 parallel electrical plasma discharge tubes with a lattice spacing of $\textit{a}$ = 50 mm. Each neighboring layer is rotated by 90 degrees and spaced $\textit{a}/2$ = 25 mm from the previous layer while offset by $\textit{a}/2$ = 25 mm to form the woodpile configuration with a total of 10 layers, as illustrated in Fig. \ref{fig:2}. The discharge tubes are held in place at four of the crystal boundaries by an acrylic scaffold to form the woodpile structure. The individually controllable alternating-current (AC) discharge tubes have a 290 mm long quartz envelope which  is 15 mm in outer diameter with a 1 mm wall thickness. The tubes are filled with argon, with added mercury to a pressure of 250 Pa. The gas temperature during nominal operation is approximately 330 K, resulting in a mercury vapor partial pressure of about 3.5 Pa. Each discharge is driven individually by an AC ballast with a peak-to-peak voltage, $V_{pp}$ = 160V. The voltage waveform of each discharge is triangular in shape resulting in a root-mean-square (RMS) voltage of $V_{RMS} = 80 / \sqrt{3}$ V. The ballasts have a variable peak current (also close to triangular) to control the discharge parameters, ranging from 24.8 mA to 111.1 mA, with a ballast frequency that decreases linearly from 55.0 kHz to 37.0 kHz for increasing peak current in the range from 24.8 mA to 51.2 mA, and a ballast frequency in the range of 32.2 kHz to 33.8 kHz for peak discharge currents from 54.4 mA to 111.2 mA. Based on previous studies\cite{Wang2015a}, where these discharge tubes were used to tune the vacancy defect of a 2D PC, we estimate that the average plasma density within the discharge is $n_{e}(cm^{-3}) \approx 5 \times 10^{9}I_{p}(mA)$.

We use the ANSYS commercial software package referred to as the High Frequency Electromagnetic Field Simulator (ANSYS HFSS 16) to compute the electromagnetic fields and transmission spectra. The simulations are carried out in 3D. The plasma columns are modeled with a frequency dependent Drude dielectric constant given by

\begin{equation}
\varepsilon_{p} = 1 - \frac{\omega_{p}^{2}}{\omega(\omega+i\nu)}
\end{equation}

where $\omega$ is the EM wave frequency, $\nu$ is the electron collisional damping rate, and $\omega_p$ is the plasma frequency, given by $\omega_p$ = $\sqrt{n_e {e}^{2}/{m}_e \varepsilon_o}$ with $n_{e}$, $\textit{e}$, and $m_{e}$ the column electron number density, electron charge, and electron mass, respectively, and $\varepsilon_o$ is the vacuum permittivity. As described in our prior study \cite{Wang2015a}, to account for the radially non-uniform plasma density we assume in the simulations that the plasma electrons are distributed uniformly over a reduced diameter of 9.2 mm, resulting in a region of electron density that is twice that of the case where the electrons are distributed uniformly over the entire inner tube diameter of d = 13 mm. We assume a reasonable value of $\nu$ = 1 GHz for the electron collisional damping rate, as this value resulted in good agreement between the measured and experimental transmission in previous studies\cite{Wang2015a}. We also include the presence of the quartz tube with an assumed dielectric constant $\varepsilon$ = 3.8. As illustrated in Fig. \ref{fig:2}(a), the source antenna is placed at the the location depicted by the red rectangle with polarization (E-field) in the $\textit{z}$-direction (transverse to the first-layer plasma columns). The receiver antenna is placed at the depicted location of the black dashed-line rectangle on the back side of the PPC. Radiative conditions are imposed on all external boundaries.

In the experiments, a pair of broadband microwave horn antennas (A INFO LB-20180 2 GHz - 18 GHz) are used as the source and detector, located in the rectangles shown in Fig. \ref{fig:2}(a), spaced 30 cm apart, also with the E field polarized in the $\textit{z}$-axis. The antennas are  connected to a HP 8722D Vector Network Analyzer to measure the s-parameters, particularly $S_{21}$, which represents the EM transmission through the crystal. The measured transmission reported below was recorded with an integration time of 5 ms per frequency point in the scan, with a source power set at -5 dBm. 

\begin{figure}[htbp]
\centering
\includegraphics[width=0.9\linewidth]{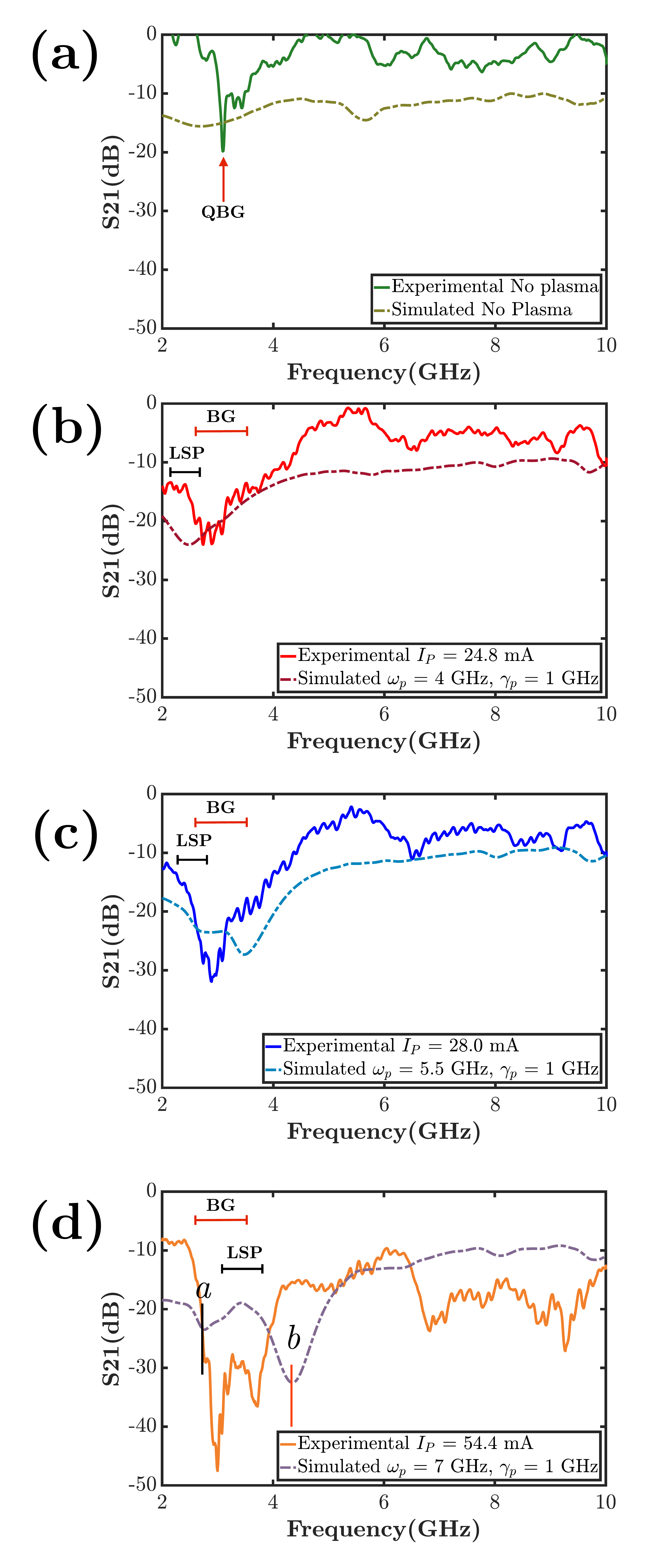}
\caption{Experimental and simulated transmission ($S_{21}$) through the 3D woodpile plasma photonic crystal. (a) Plasma off, and only quartz tubes. (b) Plasma on with discharge current $I_{P}$ = 24.8 mA, simulated $\omega_{p}$ = 4 GHz, $\gamma_{p}$ = 1 GHz. (c) Plasma on with discharge current $I_{P}$ = 28.0 mA, simulated $\omega_{p}$ = 5.5 GHz, $\gamma_{p}$ = 1 GHz. (d) plasma on with discharge current $I_{P}$ = 54.4 mA, simulated $\omega_{p}$ = 7 GHz, $\gamma_{p}$ = 1 GHz.}
\label{fig:3}
\end{figure}

\begin{figure}[htbp]
\centering
\includegraphics[width=\linewidth]{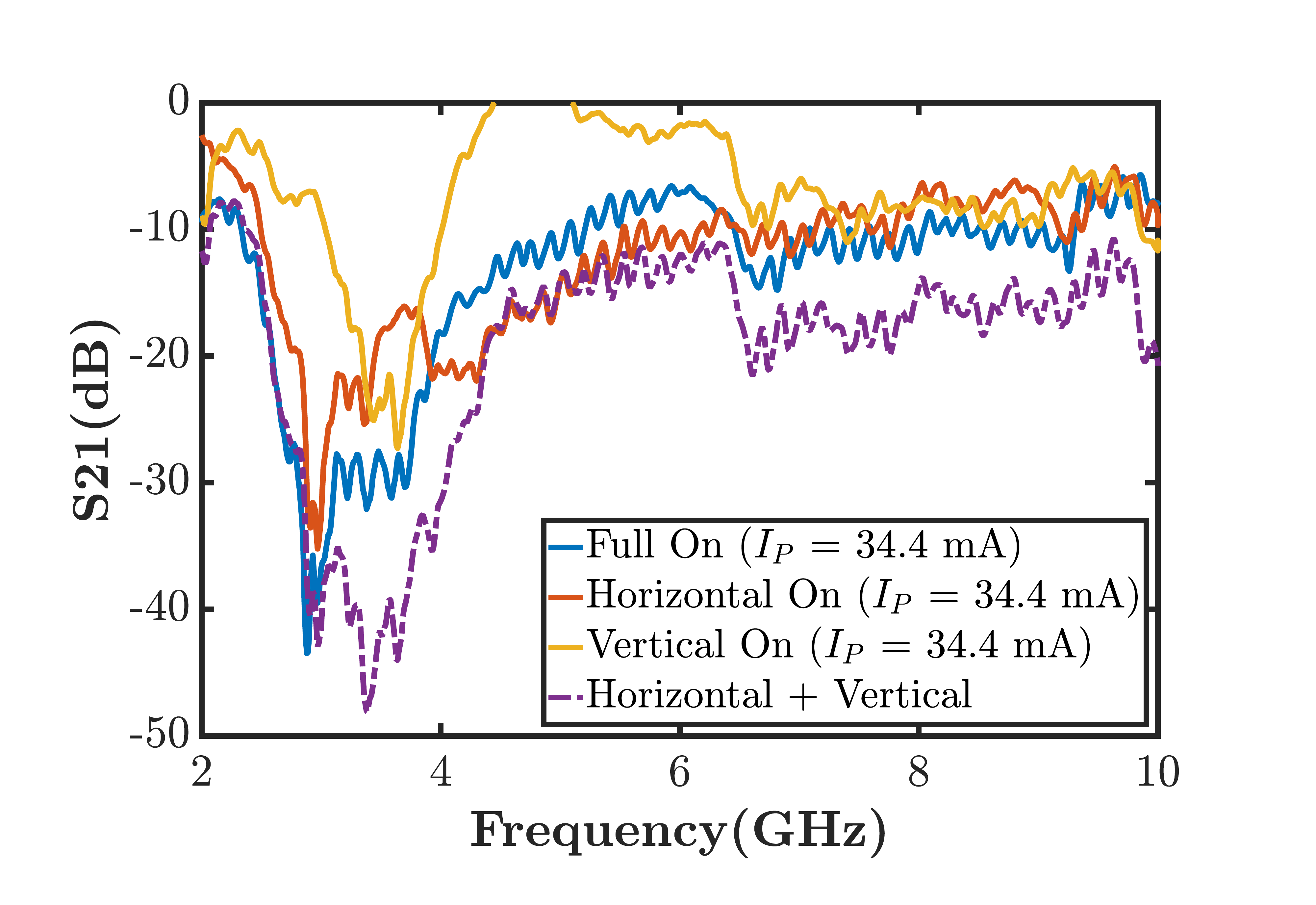}
\caption{Experimental transmission ($S_{21}$) through the 3D woodpile plasma photonic crystal for all plasma columns turned on, horizontal ($\textit{x}$-aligned) plasmas turned on, vertical ($\textit{z}$-aligned) plasmas turned on, and sum of horizontal and vertical on tube configurations with discharge current $I_{P}$ = 34.4 mA.}
\label{fig:Disp3}
\end{figure}

\begin{figure}[htbp]
\centering
\includegraphics[width=\linewidth]{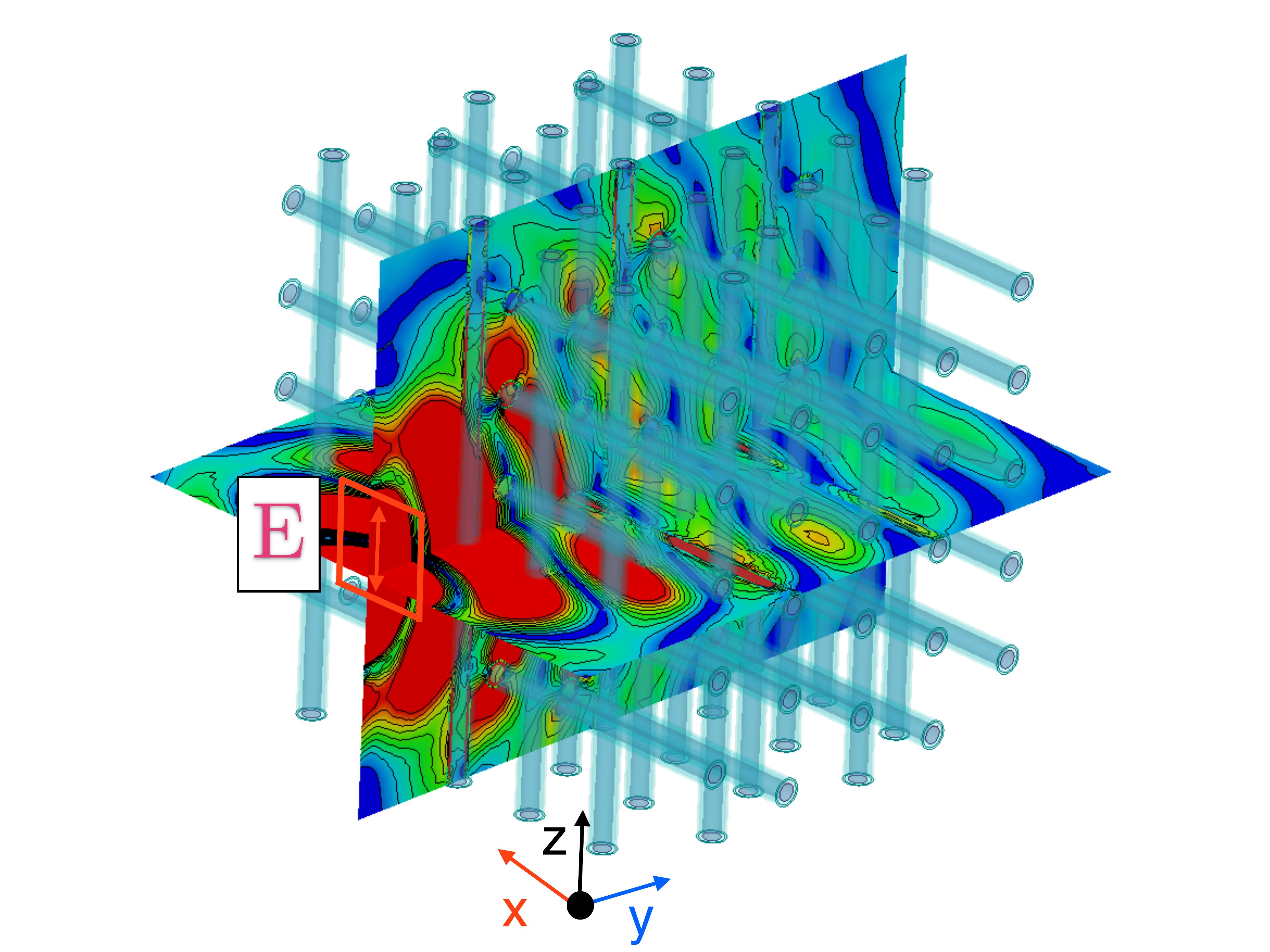}
\caption{Simulated E-fields (magnitude) for the 3D woodpile photonic crystal with $\omega_{p}$ = 7 GHz, $\gamma_{p}$ = 1 GHz at \textit{f} = 2.775 GHz. The fields are shown for the $\textit{y-z}$ and $\textit{y-x}$ planes passing through the center of the PPC.}
\label{fig:Disp4}
\end{figure}

The use of plasmas allows for localized surface plasmon (LSP) excitation at the plasma - quartz/air interface at frequencies below the plasma frequency when the E-field is in a plane perpendicular to the plasma columns. For 2D PPCs, LSPs are excited for TE polarizations only, with Bragg scattering modes expected for both polarizations. As we show below, this 3D PPC exhibits characteristics attributable to both LSP and Bragg scattering, and the attenuation is not simply the linear combination of that associated with the interwoven 5 x 5 orthogonal 2D PPCs, but instead, exhibits synergistic effects arising from the interplay between the EM activity contributed by both lattices in this woodpile configuration.

Experimental and simulated transmission spectra are shown in Fig. \ref{fig:3}. With the plasma discharges turned off (Fig. \ref{fig:3}(a)) attenuation is fairly shallow, but there are discernible band gaps attributable to the quartz throughout the spectrum, most notably near 3 GHz, with some correspondence to features in the simulations. With the plasma on and at relatively low discharge currents $I_{P}$ = 24.8 mA (Fig. \ref{fig:3}(b)) and $I_{P}$ = 28.0 mA (Fig. \ref{fig:3}(c)), the refractive index of the plasma is still close to unity and the higher frequency regions of the spectrum show little change in comparison to the plasma off case, indicating that the presence of the plasma has but a small effect on the quartz photonic bands in this region, nor do we see any new plasma-specific bands. However, there is a marked difference at frequencies below the plasma frequency. For example, in Fig. \ref{fig:3}(b), we see the emergence of what we attribute to an LSP band at a frequency below the lowest quartz Bragg gap recorded in the experiment. In cylindrical plasma columns, when the incident wavelengths are much larger than the column radius, we expect the LSP resonance to be at approximately $\omega _{LSP}=\frac{\omega_p}{\sqrt{2}}$\cite{pitarke2006theory}.  The location of the expected LSP in the figures are based on the approximate relation between $n_e$ and $I_p$ given earlier. The approximate location of the PPC Bragg resonance (BG) identified in the figures is based on the location of the quartz gap, as there is no definitive means of isolating that gap from the effect of the LSP feature. At the slightly higher discharge current (Fig. \ref{fig:3}(c)), the LSP resonance may be nearly coincident with the PPC Bragg gap and a complex spectrum emerges, suggestive of a lattice (Fano) resonance \cite{righetti2018enhanced}, with a second peak emerging near 4 GHz. The simulated spectra for these intermediate discharge current cases ($n_{e} = 2 \times 10^{11} cm^{-3}$, or $\omega_{p}$ = 4 GHz and $n_{e} = 3.8 \times 10^{11} cm^{-3}$, or $\omega_{p}$ = 5.5 GHz, respectively) capture the experimental spectra reasonably well, most notably, the emergence of a second peak (in Fig. \ref{fig:3}(c)) which may be due to the spectral splitting of coincident coupled resonators generally seen in Fano resonances. At the highest current density case (Fig. \ref{fig:3}(d)), this coupled resonance is quite pronounced as indicated by the depth of the attenuation and a splitting of the measured spectrum. For comparison, we include a simulated spectrum for a plasma frequency $\omega_{p}$ = 7 GHz ($n_{e} = 6 \times 10^{11} cm^{-3}$), although it is apparent that the predicted spectrum has the peak frequencies further apart than what is measured, in part due to an assumed plasma density that is perhaps larger than in the experiments. The shifted experimental transmission spectra compared to simulations in this region of the spectrum is not surprising, as the simulations are highly idealized with the assumption of uniform plasma columns separated from the quartz wall by vacuum. More surprising is that for this higher current density case, we see the emergence of fairly strong Bragg gaps at high frequency that are not fully captured by the simulations.   

\begin{figure}[htbp]
\centering
\includegraphics[width=0.9\linewidth]{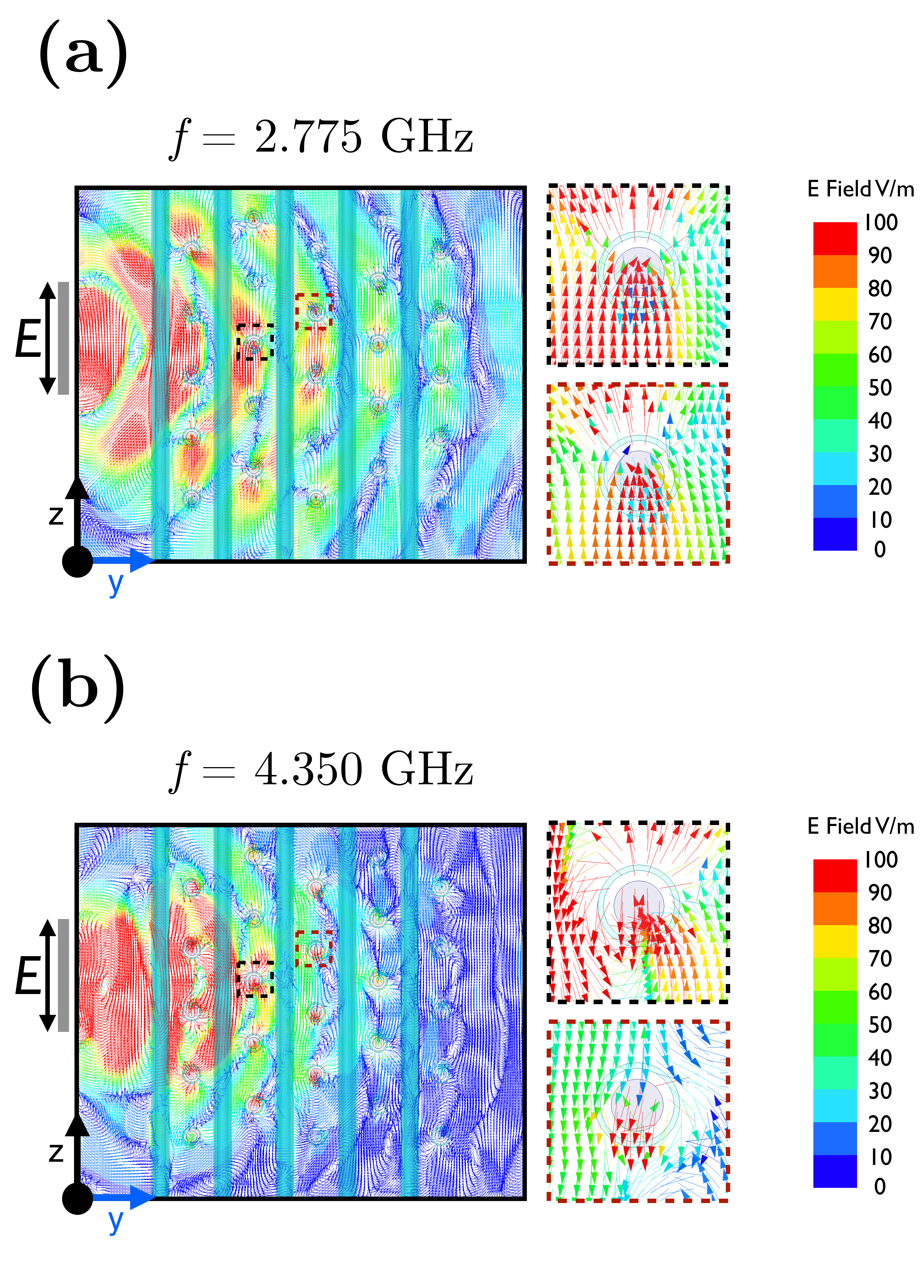}
\caption{Simulated \textit{y-z} cross section E vector field for the 3D woodpile photonic crystal with $\omega_{p}$ = 7 GHz, $\gamma_{p}$ = 1 GHz at (a) \textit{f} = 2.775 GHz (b) \textit{f} = 4.350 GHz.}
\label{fig:Disp5}
\end{figure}

We plot the transmitted spectra for a discharge current of $I_{P}$ = 34.4 mA in Fig. 4. Here we highlight three cases: (a) the case where the PPC is completely activated (the horizontal, or $\textit{x}$-aligned and vertical, or $\textit{z}$-aligned plasma tubes in Fig. 2(a)); (b) only the horizontal tubes are active; and (c) only the vertical tubes are active. For comparison to case (a), we also plot the linear sum of the transmission of (b) and (c). When only the vertical ($\textit{z}$-aligned) tubes are active we do not expect LSP excitation. The single attenuation band seen at low frequencies near 3.5 GHz is therefore the Bragg gap of the 2D hexagonal array with TM polarization. There appear to be weak Bragg gaps at frequencies $>$ 7 GHz. When only the horizontal ($\textit{x}$-aligned) tubes are active we see a split spectrum at low frequency for the 2D array with TE polarization, suggesting coupled LSP and Bragg resonances. There are no discernible features at high frequencies. Above frequencies above 3 GHz, the sum of the transmission measured for the individual 2D PPCs is much less than that measured when all plasma columns are ignited indicating that there is synergistic destructive interference between the two nested 2D arrays. Below 3 GHz, it seems that a linear summation captures the measured spectrum well, which is dominated by the 2D array with TE excitation, likely as a result of a strong LSP attenuation. At high frequencies, the attenuation is dominated by the 2D (TM) array, but the measured attenuation is still less than that predicted by the linear summation of the individual spectra. 

Some insight is gained from the simulations, particularly at lower frequencies, by examining the behavior of the predicted electric fields. To orient the reader, a 3D schematic of the PPC with superimposed E-field magnitudes (at an instant in time) is shown in Fig. 5 for $\omega_{p}$ = 7 GHz, $\gamma_{p}$ = 1 GHz at \textit{f} = 2.775 GHz. The fields are shown for the $\textit{y-z}$ and $\textit{y-x}$ planes passing through the center of the PPC. The $\textit{y-z}$ plane in Fig. 5 is the plane where the vector E-field is displayed in Fig. 6. In Fig. 6(a), the vector fields are shown for the low frequency feature labeled as ``a" in Fig. 3(d), i.e., \textit{f} = 2.775 GHz. The attenuation of the incident fields confirm the presence of Bragg interferences, and E-field phase sweeps of the region incident on two of the plasma columns with axes $\perp E$ (red and black dashed boxes) indicate the excitation of localized surface plasmons that are in phase with the incident Bragg fields. A close-up of these regions confirm that these are surface plasmons, with local E-fields that are amplified to levels above that of the background. At \textit{f} = 4.35 GHz (Fig. 6(b)), corresponding to the higher frequency feature in Fig. 3(d), again we see Bragg attenuation, and the LSPs around the same two plasma columns are even more pronounced. The presence of the LSPs at two nearby frequencies where there appear to be strong Bragg gap attenuations and the phase synchronization in time are consistent with a lattice or Fano resonance. Not apparent in these figures but evident in the simulated E-field phase sweeps are suggestions that the incident Bragg fields also couple into longitudinal surface plasmon waves that propagate along the $\textit{z}$-aligned plasma columns. 

The measurements and simulations described here, of a 3D plasma photonic crystal with a woodpile configuration, present an electromagnetic system that is rich in dynamics and an opportunity for reconfigurability that is not generally found in microwave plasma photonic crystals. We see compelling evidence for the coupling of Bragg and LSP modes through a Fano resonance. The design allows for individual control of each element in the 3D crystal. Future research will examine the propagation of waves along complex defects (associated with non-activated plasma elements) that extend in three dimensions.

\section*{Acknowledgments}
This research was supported by a Multidisciplinary University Research Initiative from the Air Force Office of Scientific Research, with Dr. Mitat Birkan as the program manager. J.A.R. acknowledges a summer fellowship provided by the Stanford University School of Engineering.

\section*{References}
\bibliography{main.bbl}

\end{document}